\documentclass[12pt]{JHEP3}

\usepackage{cite}
\usepackage{dsfont}
\usepackage{amsfonts}
\usepackage{amssymb,latexsym}
\usepackage{amsmath,amsbsy}
\usepackage{epsfig,bm}
\usepackage{graphicx,comment}
\usepackage{yfonts}

   \def\cL{{\cal L}}
\def\cM{{\cal M}}  \def\cO{{\cal O}}

\def\Im{\mathop{\rm Im}}

\newcommand{\tmfloatcontents}{}
\newlength{\tmfloatwidth}
\newcommand{\tmfloat}[5]{
  \renewcommand{\tmfloatcontents}{#4}
  \setlength{\tmfloatwidth}{\widthof{\tmfloatcontents}+1in}
  \ifthenelse{\equal{#2}{small}}
    {\ifthenelse{\lengthtest{\tmfloatwidth > \linewidth}}
      {\setlength{\tmfloatwidth}{\linewidth}}{}}
    {\setlength{\tmfloatwidth}{\linewidth}}  \begin{minipage}[#1]{\tmfloatwidth}
    \begin{center}
      \tmfloatcontents
      \captionof{#3}{#5}
    \end{center}
  \end{minipage}}

\newcommand{\sh}[1]{#1\hskip-7pt \diagup}
\newcommand{\e}{\epsilon}
\newcommand{\ee}{\epsilon^{*'}}

\newcommand{\beq}{\begin{equation}}
\newcommand{\eeq}[1]{\label{#1}\end{equation}}
\newcommand{\bea}{\begin{eqnarray}}
\newcommand{\eea}[1]{\label{#1}\end{eqnarray}}

\title{New Sum Rules from Low Energy Compton Scattering on Arbitrary Spin Target}

\author{Hovhannes~R.~Grigoryan and Massimo Porrati \\
\vspace{0.1in}

Center for Cosmology and Particle Physics\\
Department of Physics, New York University\\
4 Washington Place, New York, NY 10003, USA \\~~\\

E-mail addresses: \email{hg41@nyu.edu},
\email{massimo.porrati@nyu.edu}

\vspace{0.1in}
}

\abstract{ 
We derive two sum rules by studying the low energy Compton scattering on a target of arbitrary (nonzero) spin $j$. In the first sum rule, we consider the possibility that the intermediate state in the scattering can have spin $|j \pm 1|$ and the same mass as the target. 
The second sum rule applies if the theory at hand possesses intermediate narrow resonances with masses different from the mass of the scatterer. 
These sum rules are generalizations of the Gerasimov-Drell-Hearn-Weinberg sum rule. 
Along with the requirement of tree level unitarity, they relate different low energy couplings in the theory. 
Using these sum rules, we show that in certain cases the gyromagnetic ratio can differ from the ``natural'' value $g=2$, even at tree level, without spoiling perturbative unitarity.
These sum rules can be used as constraints applicable to all supergravity and higher-spin theories that contain particles charged under some $U(1)$ gauge field. 
In particular, applied to four dimensional $N=8$ supergravity in a spontaneously broken phase, these sum rules suggest that for the theory to have a good ultraviolet behavior, additional massive states need to be present, such as those coming from the embedding of the $N=8$ supergravity in type II superstring theory. 
We also discuss the possible implications of the sum rules for QCD in the large-$N_c$ limit.
}


\begin{document}

\section{Introduction, Summary and an Application to $N=8$ Supergravity}

Dispersion relations based on analyticity, unitarity and Lorentz invariance of scattering amplitudes can be interpreted as consistency conditions that constrain low-energy data, i.e. parameters of an effective action. Conversely, if we know the low energy effective action of a given theory, dispersion relations either constrain {\em any} of its possible  UV completions, or show that no completion exists. Beautiful examples of the latter phenomenon are given in Ref.~\cite{Adams&al}. 
In this paper we revisit topics addressed years ago by one of 
us~\cite{Ferrara:1992yc} and find their implications on the existence and properties of UV completions of low energy effective theories. In Ref.~\cite{Ferrara:1992yc} it was argued that when effective field theories of elementary particles admit a perturbative expansion (in some parametrically small dimensionless 
coupling constant), then the gyromagnetic ratio of the (weakly interacting) particles described by such theory had to be close to a 
preferred ``natural'' value: $g=2$.  This result was obtained in a Lagrangian approach. Many years earlier, 
Weinberg~\cite{Weinberg} also proposed an argument, based on a generalization of the Gerasimov-Drell-Hearn (GDH)~\cite{Gerasimov:1965et} sum rule, which similarly selected $g = 2$ as the preferred value in weakly interacting theories. 

Given a particle of mass $m$ and electric charge $e$, the GDH-Weinberg sum rule connects the gyromagnetic ratio $g$ to a dispersion integral. As usual, $g$ is defined as the ratio of the particle's magnetic moment $\mu$ to its spin $J$, so that (in the particle's rest frame):
\beq
\vec{\mu} = \frac{e g}{2m}\vec{J} \ .
\eeq{m1}

The main ingredient of the Weinberg sum rule is the low energy 
forward\footnote{Assume that the photon propagates along some $z$-direction with helicity $\lambda = \pm 1$,  and the target has a spin-$z$ projection $J_z$. For more details, see Appendix \ref{ElasticCA}.} 
Compton scattering amplitude of a photon with
energy $\omega$ and helicity $\lambda$ off a massive target of spin $J$. This amplitude, $f_{\rm scat} (\omega, \lambda)$, is a real analytic function of the  photon's energy $\omega$ away from 
the real $\omega$-axis, where cuts and poles may exist at $\omega > 0 $. The imaginary part of $f_{\rm scat}$  is given 
by the optical theorem:
\begin{align}\label{opttheorem}
{\rm Im} f_{\rm scat} (\omega,\lambda) = \frac{\omega}{4\pi}\sigma_{\rm tot}(\omega, \lambda) \ ,
\end{align}
where $\sigma_{\rm tot}$ is the total cross-section for a photon with helicity $\lambda$ and energy $\omega$. 
Define now the following function:
\begin{align}\label{minusamplitude}
&f_-(\omega^2) \equiv\frac{f_{\rm scat} (\omega,+1)-f_{\rm scat} (\omega,-1)}{2\omega} \ .
\end{align}
{\em When no intermediate (one particle) state exists in the Compton scattering, with either mass or spin different from those of the target}, then, it can be checked that (see,  e.g., Appendix \ref{ElasticCA}):
\begin{align}\label{GDH0}
&f_-(\omega^2 \to 0)  = \frac{e^2J_z}{16\pi m^2}(g-2)^2 \ .
\end{align}
Using the optical theorem (\ref{opttheorem}) and definition (\ref{minusamplitude}), we have:
\begin{align}\label{opt2}
{\rm Im}f_-(\omega^2) = \frac{1}{8\pi}\Delta\sigma (\omega) \ , \qquad
\Delta\sigma (\omega) \equiv \sigma_{\rm tot}(\omega,+1) - \sigma_{\rm tot}(\omega,-1)   \ .
\end{align}
Assuming that $f_-(\omega^2)$ vanishes when $|\omega^2| \to \infty$, one can write an {\em unsubtracted} dispersion relation:
\begin{align}\label{dispGDH}
f_-(\omega^2) = \frac{1}{4\pi^2}\int^{\infty}_0\frac{\Delta\sigma (\omega')}{\omega'^2 - \omega^2 - i \epsilon} ~\omega' d\omega' \ .
\end{align}

The validity of this assumption is far from obvious and its justification requires some additional assumptions about the UV behavior of the theory. We will argue below that unsubtracted dispersion relations hold in two important case: superstring theory and any unitary completion of $N=8$ supergravity.

When $\omega^2=0$, Eqs.(\ref{GDH0}) and (\ref{dispGDH}), impliy the 
GDH-Weinberg sum rule \cite{Weinberg,Gerasimov:1965et}:
\beq
\frac{\pi e^2 J_z}{4m^2} \left(g-2\right)^2 = \int^{\infty}_0\frac{\Delta\sigma(\omega')}{\omega'} d\omega'  \ . 
\eeq{GDH}
In weakly interacting systems, the RHS of Eq.~(\ref{GDH}) is parametrically smaller than the LHS, since in the former the final state contains at least two 
particles and is thus $O(e^4)$, while the latter is $O(e^2)$; hence $g = 2 + \cO(e^2)$.


\subsection{A Puzzle with $N=8$ Supergravity and a Possible Solution}

In truncated $N=2$ supergravity theory (where one drops the cosmological and quartic Fermionic terms from the Lagrangian given in Ref.~\cite{Freedman:1976aw}) it can be checked that the gravitino has indeed $g=2$, see also Ref.~\cite{Deser:2000dz}. 
On the other hand, in spontaneously broken $N=8$ supergravity theory~\cite{Cremmer:1979uq}, see also \cite{Andrianopoli:2002mf}, contrary to expectations, one finds that $g=1$, for both spin-1/2 and spin-3/2 fields. 
To show this, we use the dimensionally reduced action\footnote{The $N=8$ supergravity theory with global E(6) and local USp(8) invariance was constructed in five dimensions by Cremmer, Scherk and Schwarz in Ref.~\cite{Cremmer:1979uq}. Spontaneous symmetry breaking of $N=8$ supergravity theory is achieved by dimensional reduction to 4D via the Scherk-Schwarz mechanism \cite{Scherk:1978ta}.} given in Ref.~\cite{Sezgin:1981ac},  and consider only the relevant terms of the Lagrangians for spin-1/2 and spin-3/2 fields, defined as $\chi^{abc}$ and $\psi_{\mu}^a$, respectively. 
In our notations (see Appendices), and in units, $\kappa^2 = 4\pi G_N =1$, where $G_N$ is the 4D Newton's constant, these terms, written in flat space-time, unitary gauge ($\psi_5^a=0$) and near the ground state, take the following form:
\begin{align}\label{SS4d}
&\cL_{4D}(\chi) =  \frac{1}{12}\bar{\chi}^{abc}
\biggl[i\left(\sh{\partial} + 2i\cM_{abc}\sh{B}\right)
- \cM_{abc} 
+\frac{1}{4}\sigma^{\mu\nu}B_{\mu\nu} 
\biggr]\chi_{abc}  \ , \\ \label{SS4d2}
&\cL_{4D}(\psi) =  \frac{1}{2}\bar{\psi}^{a}_{\mu}
\biggl[i\gamma^{\mu\rho\nu}\left(\partial_{\nu} + 2i\cM_{a}B_{\nu}\right) - \cM_{a}\gamma^{\mu\rho}  + \frac{i}{2}\left(B^{\mu\rho} - i\gamma_5\tilde{B}^{\mu\rho}\right)\biggr]\psi_{\rho a}   \ ,
\end{align}
where $B_{\mu}$ is a graviphoton field, $\cM_{abc}$ and $\cM_a$ are the mass matrices of $\chi^{abc}$ and $\psi_{\mu}^a$ correspondingly. These spinors are $USp(8)$ tensors, and both are charged under the graviphoton, with a charge $e=2\kappa m$, where $m$ is the mass of the spinor. Notice, also that there is no $\chi$-$\psi$-$B$ mixing.

The Lagrangians above should be compared with the Dirac Lagrangian supplemented with a Pauli term, and the Lagrangian of a charged Rarita-Schwinger field with non-minimal terms \cite{Deser:2000dz}:
\begin{align}\label{DP}
&\cL_{DP} = \bar{\chi}\left[i(\sh{\partial} + ie\sh{B}) - m - \frac{e(g_{1/2}-2)}{8m}\sigma^{\mu\nu}B_{\mu\nu}\right]\chi\ , \\ \label{RS}
&\cL_{RS} = \bar{\psi}_{\mu}\left[ i\gamma^{\mu\rho\nu}(\partial_{\nu} + ie B_{\nu}) - m\gamma^{\mu\rho} + \frac{ie}{m}\left\{\frac{3}{4}\left(g_{3/2}- \frac{2}{3}\right)B^{\mu\rho} - i\alpha\gamma_5\tilde{B}^{\mu\rho}\right\}\right]\psi_{\rho} \ ,
\end{align}
where $\alpha$ is a parameter unrelated to $g_{3/2}$.\footnote{As was observed in \cite{Deser:2000dz}, the $\gamma_5$ matrix in the $\bar{\psi}_{\mu}\gamma_5 \tilde{B}^{\mu\nu}\psi_{\nu}$ term of (\ref{RS}) mixes the ``large'' and ``small'' components of $\psi_{\mu}$, and thus gives contributions of higher order in $\omega$. That is why this term does not contribute to $g_{3/2}$.} 
Finally, comparing (\ref{SS4d}--\ref{SS4d2}) with (\ref{DP}--\ref{RS}), we conclude that in case of the spontaneously broken $N=8$ supergravity: $g_{1/2} = g_{3/2}= 1$ and $\alpha =1/4$.\footnote{Similarly, comparing (\ref{RS}) with the truncated $N=2$ supergravity \cite{Freedman:1976aw}, one can deduce that: $g_{3/2} = 2$ and $\alpha = 1$. In $N=2$ supergravity with a gauged central charge \cite{Zachos:1978iw}, $g_{3/2}=2$, however, $g_{1/2} \neq 2$ or $1$.}
It appears that $g=1$ for all heavy particles in the Kaluza-Klein theory \cite{Hosoya:1983tc}. This observation was also confirmed within the string theory and on the example of D0-branes in \cite{Duff:1996bs}.

It is often the case that in supergravity theories, $e \sim m/M_P$, where $M_P$ is the Planck mass, suggesting that the charge can be made arbitrarily small. 
It is thus natural to ask: \textit{why Eq.~(\ref{GDH}) fails so miserably in case of the spontaneously broken $N=8$ supergravity theory?}
One could argue that this happens because $f_-({\omega^2})$ does not vanish sufficiently fast, when $|\omega^2| \to \infty$, since supergravity is power counting non-renormalizable. However, spontaneously broken $N=8$ supergravity can be embedded in type II superstring theory~\cite{Rohm}\footnote{Spontaneously broken  $N=4,2,1$ theories can be embedded also in heterotic string theory~\cite{KP}.}, where the condition $\lim_{\omega^2\rightarrow \infty}f_-(\omega^2)= 0$ holds, 
order-by-order in string perturbation theory. 

A more explicit and general argument supporting unsubtracted dispersion relations is
that, as argued in Refs.~\cite{gidd} (see also~\cite{Banks}), the gravitational 2-body elastic scattering amplitude $f(s,t)$ of any theory obeying Hermitian unitarity and crossing symmetry is
polynomially bound in the complex-$s$ upper half plane for fixed Mandelstam variable $t$. Hermitian unitarity implies $f(s,t)^*=f(s^*,t^*)$, thus polynomial boundedness in $s$ for the forward elastic scattering amplitude ($t=0$). Furthermore, Eq.~(\ref{bound}) suggests that the scattering amplitude $f_-(s,0)$ is not only polynomially bound, but indeed vanishes at large positive $s$ as 
$\sim 1/s$. Crossing symmetry, Hermitian unitarity, and the same argument used to prove polynomial boundedness, i.e. the Phragmen Lindel\"of theorem, then show that $f_-(s,0)$ vanishes at large $s$ in the whole complex plane.\footnote{Eq.~(\ref{SS4d}) gives a tree-level $f_-(s,0)$ that does not vanish at $s=\infty$; this fact alone shows that $N=8$ supergravity cannot be a {\em perturbatively} complete theory of gravity.}

As we will argue, the solution to this puzzle is different and it is one of the main results of this paper.
The point is that the GDH-Weinberg sum rule is modified when the Compton scattering amplitude includes intermediate one-particle states of masses $M_n \neq m$.\footnote{In this case, $\Delta\sigma$ becomes a sum of terms proportional to $\delta(\omega-\omega_n)$ plus the contribution from the continuum.} Taking into account this possibility, we find the following generalization of the GDH-Weinberg sum rule:
\begin{align}
\label{sumrule}
&{\qquad}{\qquad} \frac{e^2 J_z}{4m^2}(g-2)^2 =  [K^{\dagger},K]_{ii} +
\frac{1}{\pi}\int^{\infty}_0 \frac{\Delta \sigma}{\omega} d\omega  \ ,  \\[5pt] \nonumber
[K^{\dagger},K]_{ii} \equiv \sum_n &\left[(K_{ni})^{\dagger}K_{ni} - 
K_{in}(K_{in})^{\dagger}\right]  , {\qquad}  K_{ni} \equiv \left[2\omega_n\sqrt{m(\omega_n + m)}\right]^{-1}~\langle n |\vec{\e}_{\lambda=1}\vec{{\cal J}}|i \rangle \ , 
\end{align}
where $m$ is the mass of the target, $M_n$ is the mass of the intermediate state, $\omega_n = \frac{1}{2m}(M^2_n - m^2)$, $\vec{{\cal J}}$ is the electromagnetic current, and $\vec{\e}_{\lambda}$ is the polarization vector of the photon with helicity $\lambda$. The matrix $K_{ni}$ describes transition between states of different mass, and possibly spin.

This sum rule should be obeyed not only in string theory, but also in any Lorentz invariant, causal, unitary UV completion of $N=8$ supergravity. Because $e \sim m/M_P$,  the LHS of~(\ref{sumrule}) is $\cO(1/M_P^2)$, independent of the mass $m$, and thus is the RHS. Generically this implies that the new intermediate states have masses $M_n$ independent of $m$; these masses thus define a cutoff, below which the theory is described by $N=8$ supergravity. Type II superstring compactified on a 6-torus is a concrete example of this general situation. In this case the relation between the string mass scale $M_S\sim 1/\sqrt{\alpha'}$ and the Planck scale is $M_P=M_S^4\sqrt{V_6}/g_S$, where $V_6$ is the volume of the 6-torus, and $g_S$ is the string coupling constant.
In the perturbative regime, where $V_6M_S^6 \gg 1$ and $g_S\ll 1$, we have: $M_P\gg M_S $. 
When all radii of the torus are $\cO(1/M_S)$, $M_S$ becomes the only UV cutoff scale and $M_n=\cO(M_S)$.

The fact that new states, besides those already present in $N=8$ supergravity,\footnote{In a spontaneously broken phase, where gravitino is massive and charged \cite{Cremmer:1979uq}.} are necessary for $N=8$ to admit a UV completion has  an immediate consequence. It implies that $N=8$ supergravity is not perturbatively complete by and in itself. 
Such a statement, that is at variance with other remarkable finiteness properties of the theory (see e.g~\cite{BCJ}) warrants further discussion.
It follows from two assumptions, besides Lorentz invariance and unitarity, both satisfied by string theory but holding more generally in any perturbative regularization of gravity. The first is that the cutoff scale $\Lambda$ is parametrically smaller than $M_P$: $\Lambda= \lambda_c M_P$, where $\lambda_c \ll 1$ is the dimensionless coupling constant of the UV complete theory. The second assumption is that scattering amplitudes are regular in the limit $m\rightarrow 0$; more precisely, that the mass 
$m$ appears only with positive powers in scattering amplitudes. One can then use the fact that forward scattering depends only on the relativistic invariant $s=m^2 + 2m\omega$ to rewrite the sum rule~(\ref{sumrule}) as
\beq
\lambda^2_c \frac{\pi J_z}{4\Lambda^2}(g-2)^2 =\lambda^2_c \int_{m^2}^\infty \frac{ds}{s-m^2} \sum_n f_n[m^2,(s-m^2)/\Lambda^2]
\delta(s-M_n^2) + O(\lambda^4_c).
\eeq{sumrule2a}
Terms $\cO(\lambda^4_c)$ come from the continuum part of the total cross section, in which the final state contains at 
least two particles.  By assumption, the $m\rightarrow 0$ limit of Eq.~(\ref{sumrule2a}) is smooth, so the functions $f_n[0,x]$ are well defined, $m$-independent and dimensionless.  
In the limit $m\rightarrow 0$, dividing both sides of Eq.~(\ref{sumrule2a}) by $\lambda^2_c$, we get:
\beq
\frac{\pi J_z}{4\Lambda^2}(g-2)^2 =\int_{0}^\infty \frac{ds}{s} \sum_n f_n[0,s/\Lambda^2]\delta(s-M_n^2) + O(\lambda^2_c).
\eeq{sumrule3a}
If the LHS of~(\ref{sumrule3a}) is nonzero, then the RHS {\em must contain new states}
with masses $M_n = \cO(\Lambda)$, because, by assumption, the only massless states in the theory are those of $N=8$ supergravity.\footnote{Because of $N=8$ supersymmetry, new massless states would necessarily contain an additional massless spin two particle interacting with supergravity particles, in contradiction with the no go theorem in~\cite{P2008}.} 

One may worry that at energies above $M_P$ the cross section $\Delta \sigma$ would never be perturbative because contributions from black hole intermediate states would dominate. In general, these black holes carry mass $M$, charge $Q$ and angular momentum $J$. In such case, the cross section can be crudely approximated by the cross sectional area of a black hole~\cite{gidd}:
\begin{align}
\sigma_J \sim \pi r^2_+ \approx 4\pi G^2_N M^2-2\pi G_NQ^2 - 2\pi \frac{J^2}{M^2}
\end{align}
where $r_+ = G_N M + \sqrt{G^2_NM^2 - J^2/M^2 -G_NQ^2}$ is the outer horizon of the Kerr-Newman black hole, and we assumed $J^2 + G_N M^2 Q^2 \ll G^2_N M^4 $ or $J \ll G_NM^2$, since $Q^2 = e^2 \sim G_N m^2$ and $m \ll M$.  Clearly, the above estimate of the cross section is only applicable in case $r_+ \sim G_N M \geq 1/\Lambda$, where $\Lambda$ is a cutoff of the theory (as was mentioned above $\Lambda = \lambda_c M_P$ and $\lambda_c \ll 1$). Therefore, our estimate is valid, only when $M > \omega_c$, where $ \omega_c \equiv M_{P}\left(M_P/\Lambda\right) = \Lambda/\lambda^2_c $. In this case, the difference between the spin aligned ($J=j+1$) and anti-aligned ($J=j-1$) cross sections is:
\begin{align}\label{bound}
\Delta \sigma_{j}(\omega > \omega_c) \equiv \sigma_{j-1} -\sigma_{j+1} \sim \frac{8\pi j}{\omega^2} \ ,
\end{align}
and therefore, the dispersion integral can be divided into two parts as follows:
\begin{align}
\int_{0}^{\infty}\frac{\Delta \sigma_{j}}{\omega} d\omega  = \int_{0}^{\omega_c}\frac{\Delta \sigma_{j}}{\omega} d\omega  + \int_{\omega_c}^{\infty}\frac{\Delta \sigma_{j}}{\omega} d\omega \ , {\qquad}  \ \int_{\omega_c}^{\infty}\frac{\Delta \sigma_{j}}{\omega} d\omega \sim  \frac{4\pi j}{\omega^2_c} \sim \cO\left(\frac{\lambda^2_c}{M^2_P}\right) \ ,
\end{align}
where the second part is due to the exchange of black hole states. Since, in supergravity theories $e \sim m/M_P$, from Eq.~(\ref{GDH}) it follows that the black hole contribution to the gyromagnetic ratio is: 
\begin{align}
(g-2)^2_{\rm B.H.} \sim \cO\left(\lambda^2_c\right) \ll 1 \ .
\end{align}
If $N=8$ supergravity is a self-complete theory by itself, our argument shows that, unsurprisingly, it must unitarize at the Planck scale, in which case $\omega_c=M_P$ and the black hole contributions to the dispersive integral becomes $\cO(1)$. 

This paper also generalizes the Weinberg-GHD sum rule in another way: it allows for off-diagonal couplings in case particles of spin $j$ and $|j\pm 1|$ are degenerate in mass.
The general sum rule for Compton scattering on a target of arbitrary nonzero spin-$j$, with additional intermediate mass-degenerate states of spin $|j\pm 1|$ and possible heavy narrow resonances, can be written as: 
\begin{align}\label{sumrulesshort}
&\frac{e^2J_z}{4m^2} \left[(g_j -2)^2 - \left(j+\frac{3}{2}\right)h_{j+1/2}^2 + \left(j-\frac{1}{2}\right) h_{j-1/2}^2\right] =  [K^{\dagger},K]_{ii}  + \frac{1}{\pi}\int^{\infty}_{0} \frac{\Delta \sigma}{\omega} d\omega  \ .
\end{align}
where $g_j$ is the gyromagnetic ratio for a particle of spin $j$, and the parameters $h_{j\pm1/2}$ are couplings of the theory defined in Eq.~(\ref{hcouplings}).\footnote{Notice, that the LHS of Eq.~(\ref{sumrulesshort}) can be also written as a commutator of some generator like $K$.}

When applied to theories where all states are either proportional to a light mass scale $m$ or to a higher scale $M$ and where charges are proportional to the mass, such 
as Kaluza-Klein theories or spontaneously broken extended supergravities, this sum rule generalizes Eq.~(\ref{sumrule3a}) to
\beq 
\frac{\pi J_z}{4M^2}\left[(g_j -2)^2 - \left(j+\frac{3}{2}\right)h_{j+1/2}^2 + \left(j-\frac{1}{2}\right) h_{j-1/2}^2\right] =
\int_{0}^\infty \frac{ds}{s} \sum_n f_n[0,s/M^2]\delta(s-M_n^2) +
O(\lambda^2).
\eeq{sumrule4} 
Since magnetic dipole couplings are already taken into account by the LHS of Eq.~(\ref{sumrule4}), the dispersive integral at low $s$ contains only quadrupole or higher multipole interactions; therefore, $f_n[0,s/M^2]$ is at most $O(s^2/M^4)$ at low $s$. This implies that, if the LHS in~(\ref{sumrule4}) is 
nonzero, the RHS must contain contributions from the massive states, since the one-particle contribution to the dispersive integral due to states that become massless in the limit $m\rightarrow 0$ vanishes.

This  paper is organized as follows:  
In section 2, we describe a formalism used by Weinberg to study the low energy Compton scattering. We also study scattering on a target of spin-1/2, with possible spin-3/2 intermediate state, and deduce some generalization of the GDH-Weinberg sum rule. 
In section 3, we consider the generalization of the sum rule to Compton scattering on a target of arbitrary spin.
In section 4, we further generalize the sum rule, assuming that there are other states with masses different from the mass of the scatterer. 
In section 5, we discuss possible implications of our sum rule for nucleon 
to delta electromagnetic transitions, in the large $N_c$ limit.


\section{Compton Scattering: Weinberg's Approach}

Define the non-diagonal vertex for the emission of a single soft photon as follows:
\begin{align}
\langle \textbf{p}', s', \sigma' | {\cal J}^{\mu}(0)| \textbf{p}, s, \sigma\rangle = \Gamma_{\sigma'\sigma}^{\mu}(\textbf{p}',\textbf{p}) \ ,
\end{align}
where ${\cal J}^{\mu}$ is the conserved electromagnetic (EM) current, $| \textbf{p}, s, \sigma\rangle$ is a single-particle state with momentum $\textbf{p}$, spin-$s$ and spin $z$-component $\sigma$, as well as energy $E(\textbf{p}) = \sqrt{\textbf{p}^2+m^2}$. We adopt the following normalization:
\begin{align}
\langle \textbf{p}', s, \sigma' | \textbf{p}, s, \sigma\rangle = (2\pi)^32E_{\textbf{p}}\delta_{\sigma'\sigma}\delta^{(3)}(\textbf{p}'-\textbf{p}) \ .
\end{align}
The S-matrix, describing the Compton scattering can be written as:
\begin{align}
\langle \sigma'\textbf{p}';\lambda',\textbf{k}' |S| \sigma,\textbf{p};\lambda,\textbf{k} \rangle &=
i\ee_{\nu}(\textbf{k}',\lambda')\e_{\mu}(\textbf{k},\lambda) ~ (2\pi)^4\delta^{(4)}(p+k-p'-k')M_{\sigma'\sigma}^{\nu\mu}(\textbf{k}; \textbf{p}', \textbf{p}) \ ,
\\[5pt] 
M_{\sigma'\sigma}^{\nu\mu}(\textbf{k}; \textbf{p}', \textbf{p}) &\equiv i\int d^4x~e^{ikx} \left[
\langle \textbf{p}', s, \sigma' | T\{{\cal J}^{\nu}(0){\cal J}^{\mu}(x)\}| 
\textbf{p}, s, \sigma\rangle  + {\rm C.T.}  \right] \ ,
\end{align}
where by ${\rm C.T.}$ we mean other contact terms (seagulls), such as terms emerging from the interaction of the initial and final photon at a single point.

We will need the pole structure of $M^{\nu\mu}$ ($\sigma$, $\sigma'$ 
indices are dropped for convenience). Inserting a complete set of states between the current operators, the time-ordered product can be written as:
\begin{align}\label{amplitude}
&i\int d^4x ~ e^{ikx}\langle \textbf{p}', s, \sigma' | 
T\{{\cal J}^{\nu}(0){\cal J}^{\mu}(x) \}| \textbf{p}, s, \sigma\rangle \\ 
\nonumber  &= \int \frac{d^3p_n}{2E_n(\textbf{p}_n)}\sum_n \biggl\{\langle \textbf{p}', s, \sigma' | {\cal J}^{\nu}(0)| n \rangle \langle n| {\cal J}^{\mu}(x)| \textbf{p}, s, \sigma\rangle \frac{\delta^{(3)}(\textbf{p}_n - \textbf{p}-\textbf{k})}{E_n(\textbf{p}_n) - E(\textbf{p})- \omega - i\epsilon}  \\ \nonumber
&+\langle \textbf{p}', s, 
\sigma' | {\cal J}^{\mu}(0)| n \rangle \langle n| {\cal J}^{\nu}(x)| \textbf{p}, s, \sigma\rangle \frac{\delta^{(3)}(\textbf{p}_n - \textbf{p}'+\textbf{k})}{E_n(\textbf{p}_n)-E'(\textbf{p}') + \omega - i\epsilon} \biggr\} \\ \nonumber
&=\sum_n \biggl\{\frac{\Gamma^{\nu}(\textbf{p}',\textbf{p}+\textbf{k})\Gamma^{\mu}(\textbf{p}+\textbf{k},\textbf{p})}{2E_n(\textbf{p}+\textbf{k})[E_n(\textbf{p}+\textbf{k})-E(\textbf{p}) - \omega - i\epsilon]}  
%
+ \frac{\Gamma^{\mu}(\textbf{p}',\textbf{p}'-\textbf{k})\Gamma^{\nu}(\textbf{p}'-\textbf{k},\textbf{p})}{2E_n(\textbf{p}'-\textbf{k})[E_n(\textbf{p}'-\textbf{k})-E'(\textbf{p}') + \omega - i\epsilon]} \biggr\} \ .
\end{align}
If $|n\rangle$ is a single-particle intermediate state with the same mass $m$ as the target, then both terms in the sum above have a pole at $k^{\mu}=0$. Using Eq.(\ref{amplitude}), in the forward scattering limit, and near the $\omega=0$ pole, we get:
\begin{align}\label{forwardamplitude}
M^{\nu\mu}(\textbf{k}; \textbf{p}', \textbf{p})  \approx  &-\frac{1}{2m\omega}\biggl[\Gamma^{\nu}(\textbf{p}',\textbf{p}+\textbf{k})\Gamma^{\mu}(\textbf{p}+\textbf{k},\textbf{p}) 
-  \Gamma^{\mu}(\textbf{p}',\textbf{p}'-\textbf{k})\Gamma^{\nu}(\textbf{p}'-\textbf{k},\textbf{p})\biggr] + \ {\rm O.T.} \ ,
\end{align}
where by {\rm O.T.} we mean other terms that do not contain poles at $\omega=0$. 
As usual, the scattering amplitude will be defined as:
 \begin{align}\label{amplitudefscat}
f_{\rm scat}(\textbf{k}',\lambda'; \textbf{k}, \lambda) = \frac{1}{8\pi m}\ee_j\e_i M^{ji}(\textbf{k}',\textbf{k}, \omega)  \ .
 \end{align}
The case when $s=s'$ was considered in detail by Weinberg \cite{Weinberg}. Here, we are interested in the case when the intermediate state could be a particle of different spin (but of the same mass) as the target. To be more specific, we will consider a situation when $|s'-s|= 1$ or $0$.  

In particular, when $s=1/2$ and $s'=3/2$, the vertex function is 
[see also Eq.~(\ref{M1a})]:\footnote{This vertex clearly implies the conservation of EM current: $k_{\nu}\Gamma^{\nu}_{\sigma\sigma'} = 0 $.}
 \begin{align}\label{VF1new}
\Gamma^{\nu}_{\sigma\sigma'}(p',p'+k') &= \frac{ie \kappa_{M}}{2m^2}\epsilon^{\mu\nu\alpha\beta}p'_{\mu}k'_{\alpha} \bar{u}^{\sigma}(p')\psi^{\sigma'}_{\beta}(p'+k') \ .
 \end{align}
In case of spin-1/2 target, we take the intermediate states to be either spin-1/2 or spin-3/2 particle state. Here, we will only need to compute the part of the amplitude with spin-3/2 intermediate state, since the result in case of the spin-1/2 intermediate state is already known. Using Eqs.~(\ref{forwardamplitude}) and (\ref{amplitudefscat}) [also Eq.~(\ref{projop}) to sum over Rarita-Schwinger (RS) states] we compute:\footnote{From Eq.~(\ref{VF1new}) and current conservation: $k_{\mu}M^{\nu\mu}(\textbf{k}; \textbf{p}', \textbf{p})  = 0$, we can deduce that O.T. in Eq.~(\ref{forwardamplitude})  do not contribute to this part of the amplitude.}
\begin{align}
f_{RS}(\textbf{k}',\lambda'; \textbf{k}, \lambda)  &= \frac{i\omega e^2\kappa^2_{M}}{12\pi m^2} \left[(\vec{n}' \times \vec{\e}^{'*})\times (\vec{n} \times \vec{\e})\right]\vec{J}  \ .
\end{align}
In the forward-scattering limit, $f_{RS} = f_{RS}(\omega, \lambda)$, and we can define the following amplitude:
\begin{align}\label{gminus}
&g_{-}(\omega^2) \equiv \frac{f_{RS}(\omega, +1) -  f_{RS}(\omega, -1)}{2\omega} \ , 
\end{align}
in which case, it can be checked that:
\begin{align}\label{gminus2}
&g_{-}(\omega^2 \to 0)  = -\frac{e^2\kappa_M^2}{12\pi m^2}~J_z \ .
\end{align}
Therefore, assuming $g_-(|\omega^2| \to \infty) \to 0$, the total scattering amplitude will satisfy the following generalized unsubtracted dispersion relation:
\begin{align}\label{genGDHsr}
4\pi^2[f_-(0)+g_-(0)] &= \frac{\pi e^2}{m^2} \left(\kappa^2_p  - \frac{1}{3}\kappa_M^2\right)J_z 
= \int^{\infty}_0\frac{\sigma_{\rm tot}(\omega',+1) - \sigma_{\rm tot}(\omega',-1)}{\omega'}d\omega' \ ,
\end{align}
where $\kappa_p = (g-2)/2$ is the anomalous magnetic moment of the target (see Appendix \ref{ElasticCA}).
Eq.~(\ref{genGDHsr}) is a generalization of the GDH-Weinberg sum rule, 
when there is a spin-3/2 intermediate state in the 
Compton scattering process, that has the same mass as the spin-1/2 target. One of the consequences of this sum rule is 
that in weakly coupled theories, the gyromagnetic ratio for spin-1/2 particle can be different from its ``natural'' value $g=2$. As can 
be deduced from Eq.~(\ref{genGDHsr}), 
\begin{align}\label{gyroRSnew}
g = 2\left(1 \pm \frac{1}{\sqrt{3}}\kappa_M\right) + \cO(e^2) \ .
\end{align}
The same result can be obtained using Feynman diagrams in Appendix \ref{Feynmanway}.


\section{Generalization to Spin-$j$ Target: First Sum Rule}

The non-diagonal transition vertex for the emission of a single soft photon 
can be written as in \cite{Ferrara:1992nm}:
\begin{align}
&\Gamma^{\mu}(\textbf{k},\textbf{0}) \equiv 
\langle \textbf{k}, j', \sigma' | {\cal J}^{\mu}| \textbf{0}, j, \sigma\rangle = -2im~\epsilon^{0\mu\alpha\beta}k_{\alpha} \langle j', \sigma'|\mu_{\beta}|j,\sigma \rangle + {\cal O}(\omega^2) \ , 
\end{align} 
where $\mu_i$ is the magnetic moment and $j \neq j'$. Applying the 
Wigner-Eckart theorem, the matrix elements of the magnetic moment can be 
parametrized as follows: $\langle j,\sigma|\mu_3|j,\sigma\rangle = e g_j \sigma/2m$ and
\begin{align}
&\langle j \pm 1/2, \sigma - 1/2|\mu^{-}|j\mp 1/2,\sigma +1/2\rangle = \pm \frac{e}{2m}h_j\sqrt{(j\mp \sigma+1)(j \mp \sigma)} \ , 
\end{align} 
where $\mu^{\pm} \equiv \mu_1 \pm i\mu_2$. 
Using Eqs.~(\ref{forwardamplitude}) and (\ref{amplitudefscat}), which are valid for any spin, the non-diagonal contribution to the forward scattering amplitude on a target of arbitrary spin-$j$ ($\neq 0$) is:
\begin{align}
 &\tilde{f}_{\rm scat}= -\frac{1}{16\pi m^2\omega}\ee_{\nu}\e_{\mu}\biggl[\Gamma^{\nu \dagger}(\textbf{k},\textbf{0})\Gamma^{\mu}(\textbf{k},\textbf{0}) -  \Gamma^{\mu \dagger}(-\textbf{k},0)\Gamma^{\nu}(-\textbf{k},0)\biggr]  \\ \nonumber
&= -\frac{\omega}{4\pi}\sum_{j'\sigma'}2\langle j, \sigma|\mu_{[i}|j',\sigma'\rangle \langle j',\sigma' |\mu_{i']}|j, \sigma \rangle  (\vec{n} \times \vec{\e}^{'*})_{i} (\vec{n} \times \vec{\e})_{i'} \ .
\end{align}
Taking into account that:
\begin{align}\label{hcouplings}
&\langle j , \sigma |\mu^{-}|j\mp 1,\sigma +1\rangle = \pm \frac{e}{2m}h_{j \mp 1/2}\sqrt{(j \mp \sigma \mp 1 + 1)(j \mp \sigma \mp 1)} \ , \\ \nonumber
&\langle j , \sigma |\mu^{+}|j\pm 1,\sigma -1\rangle = \pm \frac{e}{2m}h_{j \pm 1/2}\sqrt{(j \mp \sigma \pm 1 + 1)(j \mp \sigma \pm 1)}  \ ,
\end{align} 
and applying the definition in Eq.~(\ref{minusamplitude}) for $\tilde{f}_{\rm scat}(\omega,\lambda)$, we arrive at the following result in the forward 
scattering limit:
\begin{align}
4\pi^2\tilde{f}_-(0) &= -\frac{\pi e^2}{8 m^2}J_z\left[ (2j+3)h_{j+1/2}^2 - (2j-1)h_{j-1/2}^2 \right] \ .
\end{align}
Therefore, a more general form of the sum rule can be written as:
\begin{align}\label{MainResult}
&\frac{\pi e^2J_z}{4m^2} \left[(g_j -2)^2 - \left(j+\frac{3}{2}\right)h_{j+1/2}^2 + \left(j-\frac{1}{2}\right) h_{j-1/2}^2\right]
= \int^{\infty}_0\frac{\Delta\sigma(\omega')}{\omega'}d\omega' \ .
\end{align}
As before, tree-level unitarity demands:
\begin{align}
&(g_j -2)^2 + \frac{(2j-1)}{2}h_{j-1/2}^2 = \frac{(2j+3)}{2}h_{j+1/2}^2 \ .
\end{align}
This is in agreement with the observation made, e.g., in Ref.~\cite{Ferrara:1992nm}, proposing that $g=2$, when $h_j=0$, for all $j$.
As a simple illustration, when $j=1/2$:
\begin{align}
(g_{1/2} - 2)^2 = 2h_{1}^2 \ ,
\end{align}
suggesting, $h_{1} = \sqrt{\frac{2}{3}}\kappa_M = \sqrt{2} \kappa_p$, where we recalled Eq.~(\ref{gyroRSnew}). Similarly, when $j=3/2$, 
\begin{align}
&(g_{3/2} -2)^2 - 3h_{2}^2 + h_{1}^2 = 0 \ .
\end{align}
%


\section{Generalization for Arbitrary Intermediate State: Second Sum Rule}

Consider a situation when the intermediate state in the Compton scattering process is a very narrow resonance of mass $M_n$ ($\neq m$). Then, using Eqs.~(\ref{amplitude}) and (\ref{amplitudefscat}), near each pole the forward scattering amplitude on a target at rest becomes (see also Ref.~\cite{Weinberg}):
\begin{align}
f^{(n)}(\omega \to \omega_n, \lambda) \to \frac{1}{16\pi m}\frac{\langle \textbf{0}, s, \sigma | \vec{\e}^{\ *}_{\lambda}\vec{{\cal J}}|\textbf{k}, s_n, \sigma_n \rangle \langle \textbf{k}, s_n, \sigma_n  | \vec{\e}_{\lambda}\vec{J}|\textbf{0}, s, \sigma\rangle }{\left(\sqrt{M^2_n + \omega^2} - m - \omega\right)\sqrt{M^2_n + \omega^2_n}} \ ,
\end{align}
where $\omega_n = \frac{1}{2m}(M^2_n - m^2)$, so that: $\sqrt{M^2_n + \omega^2_n} = m+\omega_n$. Now, defining:
\begin{align}
K_{ni} \equiv  
\sqrt{\frac{m}{2E_n}}~\frac{\langle \textbf{k}, s_n, \sigma_n | {\cal J}_x + 
i {\cal J}_y| \textbf{0}, s, \sigma\rangle}{M^2_n - m^2} \ , \ \ \ \ \ K_{in} \equiv  
\sqrt{\frac{m}{2E_n}}~\frac{\langle \textbf{0}, s, \sigma| {\cal J}_x + i 
{\cal J}_y|\textbf{k}, s_n, \sigma_n \rangle}{M^2_n - m^2} \ , 
\end{align}
the amplitude $f^{(n)}_{-}(\omega^2)$, determined using Eq.~(\ref{minusamplitude}), near each pole, becomes:
\begin{align}
&f^{(n)}_{-}(\omega^2) = -\frac{1}{4\pi} ~\frac{ \omega^2_n}{\omega^2 - \omega^2_n}\left[(K_{ni})^{\dagger}K_{ni} - K_{in}(K_{in})^{\dagger}\right]  \ .
\end{align}
For the total forward scattering amplitude, $f^{\rm tot}_-(\omega^2)$, with 
an appropriate contour of integration,  we will have:
\begin{align}\label{Cauchyint}
\frac{1}{2\pi i}\int_C \frac{f^{\rm tot}_{-}(\omega'^2)}{\omega'^2 - \omega^2}d\omega'^2  &= \frac{1}{\pi}\int^{\infty}_0 \frac{\Im f^{\rm tot}_-(\omega'^2)}{\omega'^2 - \omega^2 - i\e} d\omega'^2  \ .
\end{align}
In the case $\omega^2=0$ and when the integration contour encircles all the 
particle poles we obtain: 
\begin{align}\label{sumrule21}
&4\pi [f_{-}(0)+g_-(0)] - [K^{\dagger},K]_{ii} = \frac{1}{\pi}\int^{\infty}_0 \frac{\Delta \sigma(\omega')}{\omega'} d\omega'    \ , \\[5pt] \nonumber
&[K^{\dagger},K]_{ii} \equiv \sum_n \left[(K_{ni})^{\dagger}K_{ni} - K_{in}(K_{in})^{\dagger}\right]^{\omega^2_n}_0  \ .
\end{align}
It is also interesting to consider the case when $\omega^2 = \omega^2_k$ in  Eq.~(\ref{Cauchyint}). Then one obtains a relation between the derivatives of the 
matrix elements $K_{ni}$ and the dispersion integral. Notice that $f_-(0)$ and $g_-(0)$ can be expressed in terms of commutator of some generators like $K_{ni}$, in which case the left hand side of (\ref{sumrule21}) would take a more compact form.

Summarizing, when all other intermediate states in the Compton scattering have masses (and spins) different from the mass (or spin) of the scatterer, we have 
the second sum rule: 
\begin{align}\label{sumrule2}
\frac{e^2 J_z}{4m^2}(g-2)^2 &= \frac{1}{\pi}\int^{\infty}_0 \frac{\Delta \sigma}{\omega} d\omega  + [K^{\dagger},K]_{ii} \geq 0 \ .
\end{align}
Since the left hand side of Eq.~(\ref{sumrule2}) is always non negative, so should be the right hand side. 
Ignoring the integral,  we can rewrite Eq.~(\ref{sumrule2}) at leading order 
in $e^2$, as follows:
\begin{align}
g-2 &=  \pm \frac{2m}{e}\sqrt{\frac{[K^{\dagger},K]_{ii}}{J_z}} \ .
\end{align}
This means that the gyromagnetic ratio may receive corrections even at tree level, if $[K^{\dagger},K]_{ii} \geq 0$.
In supergravity theories, $e \sim m/M_P$, and we expect $g-2$ to be independent on $m$, therefore,
$[K^{\dagger},K]_{ii} \sim e^2/m^2$,
suggesting that $g-2 \sim \cO(1)$, as for spontaneously broken $N=8$ supergravity \cite{Cremmer:1979uq}. 

Finally, allowing for an intermediate state, in a Compton scattering off of a target of spin $j$, to be any narrow resonance of arbitrary mass and spin, we arrive at the most general sum rule:
\begin{align}\label{combinedsumrule}
&\frac{e^2J_z}{4m^2} \left[(g_j -2)^2 - \left(j+\frac{3}{2}\right)h_{j+1/2}^2 + \left(j-\frac{1}{2}\right) h_{j-1/2}^2\right] = [K^{\dagger},K]_{ii} + \frac{1}{\pi}\int^{\infty}_{0} \frac{\Delta \sigma}{\omega} d\omega  \ .
\end{align}
This sum rule can be extended by taking into account other global charges of the theory at hand. We will not consider such an extension here, since it depends on the specific form of the theory. Instead, below we will study a particular case, namely the example of the $N \to \Delta$ magnetic transition.


\section{Some Applications to QCD}

Application of Eq.(\ref{genGDHsr}) or (\ref{MainResult}) for $j=1/2$, directly to Compton scattering on a nucleon ($N(938)$) with a possible delta ($\Delta(1232)$) intermediate state would be wrong for two main reasons. First of all, the mass difference $\delta$ between $\Delta$ and $N$ is around $300 \ {\rm MeV}$ and the forward scattering amplitude, in the $\omega \to 0 $ limit, would behave as $\omega^2/(\delta - \omega) \sim {\cal O}(\omega^2)$. It is only when $\delta \to 0$ that the scattering amplitude becomes of order ${\cal O}(\omega)$ and contributes comparably to the forward amplitude. The second reason is that these baryons are strongly coupled systems and loop corrections may be significant. 
However, we can still make a formal use of the first sum rule if we work in the limit when $N_c \to \infty$, in which case $\delta \sim {\cal O}(1/N_c)$ (see, e.g. Refs.~\cite{Witten,ANW}). We also need to take $e \to 0$ and $N_c \to \infty$ limits in such a way that if $\Delta \sigma \sim N^k_c$, then $e^2N^{k+2}_c \ll 1$, and the dispersive integral can be safely ignored.

Comparing Eqs.(\ref{NDlagrangian}) and (\ref{M1a}) one can deduce that:
\begin{align}\label{transmom}
\mu_{p\to \Delta^+} = g_M(0) = \sqrt{\frac{2}{3}}~\kappa_M = \pm\sqrt{2} \kappa_p \ ,
\end{align}
where we work in units of nuclear Bohr magneton, and we took into account 
the appropriate isospin factor $T^3$ corresponding to the 
$p \to \Delta^+$ transition. Taking the experimental value of the proton's anomalous magnetic moment, $\kappa_p \approx 1.79$, we will obtain $\mu_{p\to \Delta^+} \approx \pm 2.54$. Unfortunately, nothing can be said about the sign, since our sum rule relates the squares of the couplings. Nevertheless, in absolute value, our result for $\mu_{p\to \Delta^+}$ is not very far from similar ones,
obtained within the framework of other models, like the Skyrme model \cite{ANW}, and holographic QCD \cite{GH09}, which respectively give the values $2.3$ and $2.58$ (the experimental value is: $\mu_{p\to \Delta^+} =3.46 \pm 0.03$ \cite{Tiator:2000iy}).

In all models of baryons in the large-$N_c$ limit, baryons are finite size objects, whose sizes ($R$) do not scale with $N_c$, while their masses ($M$) do. As 
in the case of the GDH-Weinberg sum rule \cite{Brodsky:1968ea}, in the zero radius limit, when $M R \to 0$, we expect the magnetic transition moment to 
approach its canonical value (\ref{transmom}).  However, in the Skyrme model $R$ is fixed and $M R \sim {\cal O}(N_c) $. This is not surprising, since the Lagrangian of the Skyrme model \cite{ANW} is known to behave badly at high energies. 

Now, we want to apply the second sum rule to the same system. In this case, we take the physical masses of $N$ and $\Delta$ and only assume that these are narrow resonances. Although in this case the scattering amplitude with $\Delta$ intermediate state vanishes when $\omega \to 0$, it contributes to the dispersion integral, when $\omega \to \omega_n$. Employing Ref.~\cite{Pascalutsa:2002pi}, we obtain:
\begin{align}
&\langle \textbf{k}, 3/2, \sigma' | \vec{\e}_{\lambda}\vec{{\cal J}}| \textbf{0}, 1/2, \sigma\rangle = \sqrt{\frac{3m_N}{2M_{\Delta}}}~\frac{ie \mu_{p\to\Delta^+} }{(m_N+M_{\Delta})}\omega ~\bar{\psi}^{\sigma'}_r(\textbf{k})u^{\sigma}(\textbf{0})~(\vec{n} \times \vec{\epsilon}_{\lambda})_r \ .
\end{align}
After straightforward computations, the second sum rule gives:
\begin{align}
\kappa_p 
=  \pm 2 \frac{m_N^2}{(m_N+M_{\Delta})}~\frac{\mu_{p\to\Delta^+}}{\sqrt{m^2_N + M^2_{\Delta}}} \ .
\end{align}
Numerically, $\mu_{p\to\Delta^+} \approx \pm 1.91\kappa_p$; therefore, taking $\kappa_p=1.79$, we obtain $\mu_{p\to\Delta^+} \approx \pm 3.42$. In absolute value this result is coincidentally close to the experimental value: $\mu_{p\to \Delta^+} =3.46 \pm 0.03$ \cite{Tiator:2000iy}.

Consider another example (also relevant for large-$N_c$ QCD), when the target has spin-1/2 and the intermediate particle is an excited state with the same spin. Then, using Ref.~\cite{Devenish:1975jd}, the general form of the transition matrix element can be written as:
\begin{align}
&\sqrt{\frac{m}{2E_n}}~\frac{\langle \textbf{k}, 1/2^*,\sigma' |
{\cal J}_x + i \lambda {\cal J}_y| \textbf{0}, 1/2,\sigma\rangle}{M^2_n - m^2}
= 2e G_{*} \omega~\frac{m}{(M_n-m)}\sqrt{\frac{mM_n}{M_n^2+m^2}}~(\sigma_1 + i\lambda \sigma_2)_{\sigma\sigma'}\ ,  
\end{align}
where $G_*$ is some dimensionful coupling. Direct computations show that $[K^{\dagger},K]_{ii} < 0$, which means that for this theory with excited states to be unitary at tree level, we need $G_*=0$. However, this conclusion might change if we include intermediate states with spin-3/2.

 

\acknowledgments

This work is supported by NSF grant PHY-0758032 and by ERC Advanced Investigator Grant n.226455 {\em Supersymmetry, Quantum Gravity and Gauge Fields (Superfields)}. We would like to thank Sergio Ferrara, Gregory Gabadadze and Sergei Dubovsky for interesting discussions.


\appendix

\renewcommand{\theequation}{A\arabic{equation}}
  \setcounter{equation}{0}

\section{Low Energy Compton Amplitude for a Spin-1/2 Target}
\label{ElasticCA}

Consider the Compton scattering process: 
\begin{align}\label{process}
N\{p,\sigma\} + \gamma\{k,\e_{\mu}(k)\} \to N\{p',\sigma'\} + \gamma\{k',\e'_{\nu}(k')\}  \ , 
\end{align}
where $p$ and $p'$ are initial and final 4-momenta of scatterer (which is a spin-1/2 particle), $\sigma$ and $\sigma'$ are projections of initial and final spin along the $z$-direction. Analogously, $k$, $\e_{\mu}(k)$ and $k'$, $\e'_{\nu}(k')$ are the 4-momenta and polarizations of initial and final photons. We take the external particles to be on-shell, that is: $k_0 = |\textbf{k}|$, $k_0' = |\textbf{k}'|$, $p_0 = \sqrt{\textbf{p}^2 + m^2}$ and $p'_0 = \sqrt{\textbf{p}'^2 + m^2}$. 
The scattering amplitude corresponding to process in (\ref{process}) is:
\begin{align}\nonumber
{\cal M} &= -e^2\bar{u}(p',\sigma')\biggl\{
\Gamma^{\nu}(p',p+k)\ee_{\nu}(k')\frac{\sh{p} + \sh{k} + m}{(p+k)^2 - m^2}\Gamma^{\mu}(p+k,p)\e_{\mu}(k) \\ 
&+ \Gamma^{\mu}(p',p-k')\e_{\mu}(k)\frac{\sh{p} - \sh{k}' + m}{(p-k')^2 - m^2}\Gamma^{\nu}(p-k',p)\ee_{\nu}(k')
\biggr\}u(p,\sigma)  \ , \\[8pt] 
&{\qquad}{\qquad} \Gamma^{\mu}(p_2,p_1) \equiv \gamma^{\mu}F_D(q^2) + \frac{i\sigma^{\mu\nu}}{2m}q_{\nu}F_P(q^2) \ ,
\end{align}
where $e$ is the electric charge of scatterer, $m$ is its mass, and $q = p_2 -p_1$. Using the notations of Peskin \& Schroeder, $\sh{a} \equiv a_{\mu}\gamma^{\mu}$, $\sigma^{\mu\nu} =i[\gamma^{\mu},\gamma^{\nu}]/2$. Here, $F_D(q^2)$ and $F_P(q^2)$ are Dirac and Pauli form factors, which for $q^2=0$ are: $F_D(0) = 1$ and $F_P(0)=\kappa_p$, where $e\kappa_p/(2m)$ is the anomalous magnetic moment of the scatterer. The latter arises 
from the Pauli Lagrangian:
 \begin{align}\label{Pauli}
 {\cal L}_{\rm P} = -\frac{e\kappa_p}{4m}~\bar{u}~\sigma^{\mu\nu}u ~F_{\mu\nu} \ .
 \end{align}

In what follows, we will adopt a `gauge' in which the initial and final photon are transverely polarized in the laboratory frame. That is, we choose:
\begin{align}
\e(k) \cdot k = \ee(k') \cdot k' = \e(k) \cdot p = \ee(k') \cdot p = 0 \ ,
\end{align}
where $p=(m,\textbf{0})$, implying that $\e^0 = \e'^0 = 0$ and 
$\vec{\e}~\vec{k} = \vec{\e}^{~'}\vec{k}' = 0$.  
We also adopt the following normalization: $\e(k) \cdot \e^*(k) = \e'(k')\cdot \ee(k') = -1$. 
Using the Dirac equation: $(\sh{p}-m)u(p)=0$, and after some simplifications, we can rewrite the amplitude as follows:
\begin{align}
{\cal M} &=\frac{e^2\mu}{2m}\bar{u}(p',\sigma')\biggl[ \left(\sh{\e}' + \mu' \sh{\e}'\sh{k}'\right)\left(\frac{\sh{\e}\sh{k}}{\omega} +  \frac{1-\mu}{\mu}\sh{\e}\right)
+  \left(\sh{\e} -  \mu' \sh{\e}\sh{k}\right)\left(\frac{\sh{\e}'\sh{k}'}{\omega'} + \frac{1-\mu}{\mu}\sh{\e}'\right)\biggr] u(p,\sigma) \ , 
\end{align}
where $\omega \equiv pk/m $, $\omega' \equiv pk'/m$, $\sh{\e}' \equiv \gamma^{\mu}\ee_{\mu}$, $\mu' \equiv \kappa_p/(2m)$ and $\mu = 1 + 2m\mu'$ is the magnetic moment.
Since we work in the frame where $p_0=m$ and $\textbf{p}=\textbf{0}$, we have: $\omega = k_0$ and $\omega' = k_0'$. The initial state is $u^T(p,\sigma) = \sqrt{m}\left\{\xi,\xi\right\}$, where $\xi$ is a spinor such that $\xi^{\dagger}\xi = 1$. Similarly, for $|\textbf{p}'| \ll m $, 
\begin{align}
\bar{u}(p',\sigma') = \sqrt{m}\left\{\xi'^{\dagger}\left(1 + \frac{1}{2m}\vec{\sigma} \textbf{p}'\right), \xi'^{\dagger}\left( 1 - \frac{1}{2m}\vec{\sigma} \textbf{p}'\right)\right\} \ .
\end{align}

Taking $\omega = \omega'$ and defining $\vec{n} \equiv \vec{k}/\omega$ and $\vec{n}' \equiv \vec{k}'/\omega'$, we can perform direct matrix and vector multiplications to arrive to the final result. 
After tedious but straightforward calculations, the answer can be written in a familiar form \cite{Low:1954kd}:
\begin{align}
f_{\rm scat}  &= \frac{e^2}{4\pi m}\biggl\{-\vec{\e}^{'*}\vec{\e} +\frac{i\kappa_p}{m}\omega ~\vec{J}~(\vec{\e}^{'*} \times \vec{\e}) - \frac{i\mu^2}{m}\omega~\vec{J}~\left[(\vec{n}' \times \vec{\e}^{'*}) \times (\vec{n} \times \vec{\e})\right] \\ \nonumber
&- \frac{i\mu}{2m}~\omega~\vec{J}~\left[
\left\{\vec{n}(\vec{n}\times \vec{\e}) + (\vec{n}\times \vec{\e})\vec{n}\right\}\vec{\e}^{'*} 
-
\{\vec{n}'(\vec{n}'\times \vec{\e}^{'*}) + (\vec{n}'\times \vec{\e}^{'*})\vec{n}'\}\vec{\e}
\right]\biggr\} \ ,
\end{align}
where $f_{\rm scat} \equiv{\cal M}/(8\pi m) $ and $\vec{J}\equiv \delta_{\sigma'\sigma} \xi^{\dagger}_{\sigma} \vec{\sigma} \xi_{\sigma}/2$, with $\vec{J}$ being the spin of the scatterer (when $\sigma=\sigma'$).

\renewcommand{\theequation}{B\arabic{equation}}
  \setcounter{equation}{0}

\section{General Properties of Rarita-Schwinger Field}


The Lagrangian for a free massive spin-3/2, Rarita-Schwinger (RS) field, $\psi_{\mu}$, can be written as:
\begin{align}
{\cal L}_{\rm RS} = \bar{\psi}_{\mu}\left(i \gamma^{\mu\nu\rho}\partial_{\rho} - m \gamma^{\mu\nu} \right)\psi_{\nu} \  ,
\end{align}
where $\gamma^{\mu\nu} \equiv \gamma^{[\mu}\gamma^{\nu]}$ and $\gamma^{\mu\nu\rho} \equiv \gamma^{[\mu}\gamma^{\nu}\gamma^{\rho]}$. The equations of motion that follow from this Lagrangian can be equivalently written as Dirac equations along with the transversality and tracelessness constraints:
\begin{align}
\left(i\sh{\partial} - m\right)\psi_{\mu} = 0 \ , \ \ \ \partial^{\mu}\psi_{\mu} = 0 \ , \ \ \ \gamma^{\mu}\psi_{\mu} = 0 \ .
\end{align}
These constraints guarantee that among $16$ independent components of RS field only  $2s+1 = 4$ physical degrees of freedom will propagate. 

The wave function for the RS field can be written as a product of massive spin-1 and spin-1/2 polarizations: $e_{\mu}(\vec{p},\textit{n})$ with $\textit{n}=-1,0,1$, and $u(\vec{p},\sigma)$ with ($\sigma=\pm 1/2$), respectively. More specifically (see, e.g. Refs.~\cite{Auvil}):
\begin{align}\label{wavefunc}
\psi_{\mu}(\vec{p},r) &= \sum_{\sigma,\textit{n}} \langle\left(\frac{1}{2},\sigma\right)\left(1,\textit{n}\right)|\left(\frac{3}{2},r\right)\rangle~u(\vec{p},\sigma)~e_{\mu}(\vec{p},\textit{n}) \ ,
\end{align}
where $u(\vec{p},\sigma)$ satisfies equations: $(\sh{p}-m)u(\vec{p},\sigma)=0$ and $\hat{\sigma}_z u(\vec{p},\sigma) = \sigma u(\vec{p},\sigma) $. 
The polarization vectors satisfy the following normalization and transversality conditions: 
\begin{align}
&e^*_{\mu}(\vec{p},\textit{n})e^{\mu}(\vec{p},\textit{n}') = -\delta_{\textit{n}\textit{n}'} \ , {\qquad}
p^{\mu}e_{\mu}(\vec{p},\textit{n}) = p^{\mu}e^*_{\mu}(\vec{p},\textit{n}) = 0 \ .
\end{align}
Substituting the values of the Clebsch-Gordan coefficients in Eq.~(\ref{wavefunc}), that are proportional to $\delta_{r,\textit{n}+\sigma}$, we will get:  
\begin{align}
\psi_{\alpha}^{\pm3/2} &= e_{\alpha}^{\pm 1}u^{\pm 1/2} \ , {\qquad}
\psi_{\alpha}^{\pm 1/2} = \frac{1}{\sqrt{3}}\left(e_{\alpha}^{\pm 1}u^{\mp 1/2} + \sqrt{2}e_{\alpha}^0 u^{\pm 1/2}\right) \ , 
\end{align}
where $\psi_{\alpha}(\vec{p},r) \equiv \psi_{\alpha}^{r}$, $u(\vec{p},\sigma) \equiv u^{\sigma}$ and $e_{\mu}(\vec{p},\textit{n}) \equiv e_{\mu}^\textit{n}$. These solutions satisfy the Dirac equation, as well as transversality and tracelessness constraints. Moreover, the RS states are normalized as:
\begin{align}
&\bar{\psi}_{\alpha}(\vec{p},r)\psi^{\alpha}(\vec{p},r') = -2m \delta_{rr'} \ , {\qquad}
\bar{\psi}_{\alpha}(\vec{p},r)\gamma_{\mu}\psi^{\alpha}(\vec{p},r') = -2p_{\mu} \delta_{rr'} \ .
\end{align}
In what follows it would be useful to note that:
\begin{align}\label{projop}
P_{\mu\nu}(p) &\equiv \sum_{r}\psi_{\nu}(\vec{p},r)\bar{\psi}_{\mu}(\vec{p},r) = -(\sh{p} + m)\Pi_{\mu\nu}(p) \ ,  \\ \nonumber
\Pi_{\mu\nu}(p) &\equiv \left(g_{\mu\nu} -\frac{p_{\mu}p_{\nu}}{m^2}\right) - \frac{1}
{3}\left(\gamma_{\mu} -\frac{p_{\mu}}{m}\right)\left(\gamma_{\nu} +\frac{p_{\nu}}{m}\right) \ .
\end{align}

In the chiral representation of the $\gamma$-matrices (as in Peskin \& Schroeder), it can be checked that, when $p = (m,0,0,0)$, the solutions for $u^{\sigma}$ and $e_{\mu}^\textit{n}$ are:
\begin{align}
&u^{-1/2} = \sqrt{m}(0,1,0,1)^T \ , \ \ \ \ u^{+1/2} = \sqrt{m}(1,0,1,0)^T \ , \\ \nonumber
&e^+_{\mu} = \frac{1}{\sqrt{2}}(0,1,i,0) \ , \ \ \ e^-_{\mu} = -\frac{1}{\sqrt{2}}(0,1,-i,0) \ , \ \ \ e^0_{\mu} = (0,0,0,-1) \ .
\end{align}
%


Writing the quantized RS field as, 
\begin{align}
&\Psi_{\mu}(x) =  \int \frac{d^3p}{(2\pi)^3}\frac{1}{2E_{\textbf{p}}}\sum_{\lambda}\left\{ 
e^{-ipx}\psi_{\mu}(\vec{p},\lambda)a_{\vec{p},\lambda} +  e^{ipx}\psi^C_{\mu}(\vec{p},\lambda)a^{\dagger}_{\vec{p},\lambda} 
\right\} \ , \\
&\{a_{\vec{p},\lambda}, a^{\dagger}_{\vec{p}',\lambda'} \} = (2\pi)^32E_{\textbf{p}}\delta_{\lambda\lambda'}\delta(\textbf{p}-\textbf{p}') \ ,   
{\qquad}
\{a_{\vec{p},\lambda}, a_{\vec{p}',\lambda'} \} = \{a^{\dagger}_{\vec{p},\lambda}, a^{\dagger}_{\vec{p}',\lambda'} \} =  0 \ ,
\end{align}
and using Eq.~(\ref{projop}), it can be deduced that:
\begin{align}
&\langle 0| \Psi_{\nu}(x) \bar{\Psi}_{\mu}(y) |0 \rangle = \int \frac{d^3p}{(2\pi)^3}\frac{1}{2E_{\textbf{p}}}\sum_{\lambda}\psi_{\nu}(\vec{p},\lambda)\bar{\psi}_{\mu}(\vec{p},\lambda) e^{-ip(x-y)} \\ \nonumber
&=-(i\sh{\partial}_x+m)\Pi_{\mu\nu}(i\partial_x) \int \frac{d^3p}{(2\pi)^3}\frac{1}{2E_{\textbf{p}}}e^{-ip(x-y)} \ , \\ 
&\langle 0| \bar{\Psi}_{\mu}(y)\Psi_{\nu}(x)  |0 \rangle = \int \frac{d^3p}{(2\pi)^3}\frac{1}{2E_{\textbf{p}}}\sum_{\lambda}\psi^C_{\nu}(\vec{p},\lambda)\bar{\psi}^C_{\mu}(\vec{p},\lambda) e^{-ip(y-x)} \\ \nonumber
&=(i\sh{\partial}_x+m)\Pi_{\mu\nu}(i\partial_x) \int \frac{d^3p}{(2\pi)^3}\frac{1}{2E_{\textbf{p}}}e^{-ip(y-x)} \ .
\end{align}
Since the Feynman propagator is defined as $S^F_{\mu\nu}(x-y) \equiv \langle 0 |T\Psi_{\nu}(x) \bar{\Psi}_{\mu}(y)  | 0 \rangle $, we have:
\begin{align}
&S^F_{\mu\nu}(x-y) = -(i\sh{\partial}_x+m)\Pi_{\mu\nu}(i\partial_x) D_F(x-y) \ , \\ 
&D_F(x-y) \equiv \int \frac{d^4p}{(2\pi)^4}~\frac{i}{p^2-m^2 +i\epsilon} e^{-ip(x-y)} \ ,
\end{align}
where $D_F(x-y)$ is the Feynman propagator of a free scalar field. More explicitly, 
\begin{align}\label{propagRS}
S^F_{\alpha\beta}(p) &= -i\frac{\left(\sh{p} + m\right)}{p^2-m^2 + i\epsilon}\left[g_{\alpha\beta}  - \frac{1}{3}\gamma_{\alpha}\gamma_{\beta} - \frac{2p_{\alpha}p_{\beta}}{3m^2}
-\frac{\left(\gamma_{\alpha}p_{\beta}- \gamma_{\beta}p_{\alpha}\right)}{3m}\right]\ .
\end{align}

\renewcommand{\theequation}{C\arabic{equation}}
  \setcounter{equation}{0}

\section{First Sum Rule for Spin-1/2 Target: Alternative Approach}\label{Feynmanway}

The most general form of the vertex function, describing interactions between photon, spin-1/2 and spin-3/2 particles of the same mass, following Jones and Scadron \cite{Jones:1972ky} can be written in terms of magnetic dipole ($g_M$), electric quadrupole ($g_E$), and Coulomb quadrupole ($g_C$) form factors.
This interaction vertex effectively emerges from the following Lagrangian (see, e.g. \cite{Pascalutsa:2002pi}):
 \begin{align}\label{NDlagrangian}
 {\cal L}_{\rm int} = \frac{3ie}{4m^2}~\bar{u}T^3&\biggl[g_M \left(\partial_{\mu}\psi_{\nu}\right) \tilde{F}^{\mu\nu} + i g_E \gamma_5 \left(\partial_{\mu}\psi_{\nu}\right)F^{\mu\nu} -\frac{2g_C}{m}\gamma_5\gamma^{\alpha}\partial_{[\alpha}\psi_{\nu]}\partial_{\mu}F^{\mu\nu} \biggr] + {\rm h.c.} \ ,
 \end{align}
where $\tilde{F}^{\mu\nu} = \epsilon^{\mu\nu\alpha\beta}F_{\alpha\beta}/2$ and $T^3$ is an operator due to additional internal degrees of freedom (such as global or isospin charge) that the fields could carry. 

In the soft momentum transfer (or near forward) limit that we are interested in, only the first term will matter.
We will rewrite the Lagrangian describing the magnetic dipole transition as:
 \begin{align}\label{M1a}
 {\cal L}^{M1}_{\rm int} &= \frac{ie\kappa_{M}}{2m^2}~\bar{u}\left(\partial_{\mu}\psi_{\nu}\right) \tilde{F}^{\mu\nu}  + {\rm h.c.}  \ .
 \end{align}
The interaction vertex describing the dominant magnetic dipole $3/2 \to 1/2$ transition is:
 \begin{align}\label{VF1}
\Gamma^{\nu \beta}(p',p'+k') &= \frac{e \kappa_{M}}{2m^2}\epsilon^{\mu\nu\alpha\beta}p'_{\mu}k'_{\alpha}  \ , 
 \end{align}
where $p'$ is the momentum of spin-1/2 state and $k'$ is the photon momentum. Similarly, the vertex of $1/2 \to 3/2$ transition is:
 \begin{align}\label{VF2}
\Gamma^{\nu \beta}(p+k,p) &= -\frac{e\kappa_{M}}{2m^2}\epsilon^{\mu\nu\alpha\beta}p_{\mu}k_{\alpha}  \ , 
 \end{align}
where $p$ is the momentum of spin-1/2 state and $k$ is the photon momentum. 
For forward scattering, when $p_1 = p_2 = (m,0,0,0)$ and $\omega = \omega'$, we have:
 \begin{align}
 \Gamma^{\nu\beta}(p',p'+k') = -\Gamma^{\nu \beta}(p+k,p) &= \frac{e\kappa_{M}}{2m}\omega~\epsilon^{0\nu\beta k}n_{k}  \ . 
 \end{align}

Consider the scattering in~(\ref{process}), where the intermediate state is a spin-3/2 Rarita-Schwinger particle with the same mass as the scatterer. We want to find the amplitude corresponding to this process. It can be written as:
\begin{align}\nonumber
i{\cal M_{RS}} &= \bar{u}(p',\sigma')\biggl\{
\Gamma^{\nu \alpha}(p',p+k)\ee_{\nu}(k')S_{\alpha\beta}(p+k)\Gamma^{\mu \beta}(p+k,p)\e_{\mu}(k) \\ 
&+ \Gamma^{\mu \beta}(p',p-k')\e_{\mu}(k)S_{\beta\alpha}(p-k')\Gamma^{\nu \alpha}(p-k',p)\ee_{\nu}(k')
\biggr\}u(p,\sigma)  \ ,
\end{align}
where vertices are defined in Eqs.(\ref{VF1})-(\ref{VF2}), and propagator of RS field, $S_{\alpha\beta}$, is given by Eq.~(\ref{propagRS}).

Since we are interested in the near forward scattering amplitude up to order ${\cal O}(\omega)$, and using Eqs.(\ref{VF1}) and (\ref{VF2}), direct computations for $f_{RS} = {\cal M}_{RS}/(8\pi m)$ give:
\begin{align}\nonumber
f_{RS}(\omega, \lambda) &= \frac{i}{12\pi}\frac{e^2\kappa_M^2}{m^2}~\omega \left[(\vec{\e}^{'*} \times \vec{n}') \times \left(\vec{\e} \times \vec{n}\right)\right]\vec{J} = -\frac{\lambda\omega}{12\pi}\frac{e^2\kappa_M^2}{m^2}~J_z \ .
\end{align}
Finally, using the definition for $g_-(\omega^2)$, given in Eq.(\ref{gminus}), in the forward limit we have:
\begin{align}\label{gminusDiagram}
4\pi^2g_-(0)  &= -\frac{\pi e^2\kappa^2_{M}}{3 m^2}J_z \ ,
\end{align}
which is in agreement with the previous result (\ref{gminus2}), as should be expected.



\end{document}